# Introducing Partial Matching Approach in Association Rules for Better Treatment of Missing Values

SHARIQ BASHIR, SAAD RAZZAQ, UMER MAQBOOL, SONYA TAHIR,
A. RAUF BAIG
Department of Computer Science (Machine Intelligence Group)
National University of Computer and Emerging Sciences
A.K.Brohi Road, H-11/4, Islamabad.
Pakistan
shariq.bashir@nu.edu.pk,  rauf.baig@nu.edu.pk

*Abstract:* - Handling missing values in training datasets for constructing learning models or extracting useful information is considered to be an important research task in data mining and knowledge discovery in databases (KDD). In recent years, lot of techniques are proposed for imputing missing values by considering attribute relationships with missing value observation and other observations of training dataset. The main deficiency of such techniques is that, they depend upon single approach and do not combine multiple approaches, that's why they are less accurate. To improve the accuracy of missing values imputation, in this paper we introduce a novel partial matching concept in association rules mining, which shows better results as compared to full matching concept that we described in our previous work. Our imputation technique combines the partial matching concept in association rules with k-nearest neighbor approach. Since this is a hybrid technique, therefore its accuracy is much better than as compared to those techniques which depend upon single approach. To check the efficiency of our technique, we also provide detail experimental results on number of benchmark datasets which show better results as compared to previous approaches.

## 1 Introduction

Data quality is a major concern in machine learning and other correlated areas such as knowledge discovery from databases (KDD), data mining and data warehousing. As most machine learning algorithms induce knowledge strictly from data, the quality of the knowledge extracted is largely determined by the quality of the underlying data.

The quality of training data depends upon many factors [1, 5], but handling missing values is considered to be a crucial factor in overall data quality. For many real-world applications of KDD and data mining, even when there are huge amounts of data, the subset of cases with complete data may be relatively small. Training as well as testing samples have missing values. Missing data may be due to different reasons such as refusal to responds, faulty data collection instruments, data entry problems and data transmission problems. Missing data is a problem that continues to plague data analysis methods. Even as analysis methods gain sophistication, we continue to encounter missing values in fields, especially in databases with a large number of fields. The absence of information is rarely beneficial. All things being equal, more data is almost always better. In most cases, data sets attributes are not independent from each other. Thus, through the identification of relationships among attributes, missing values can be determined. Imputation is a term that denotes a procedure that replaces the missing values in a data set by some plausible values. One advantage of this approach is that the missing data treatment is independent of the learning algorithm used. This allows the user to select the most suitable imputation method for each situation.

In our previous work we proposed a novel missing value imputation approach (HMiT) using association rules mining with hybrid combination of k-nearest neighbor approach. In HMiT the missing values are first tried to impute using association rules by comparing the known attribute values of missing value observation with the antecedent part of association rules. In the case, when there is no rule exist or in the other words no attribute relationship exist in the training data against the missing value of an observation, then the missing values are imputed using k-nearest neighbor approach. In our different experimental results on real benchmark datasets, we show that this approach not only increases the prediction accuracy but also sufficiently decrease the processing time of missing values treatment.

To further improve the accuracy of missing values treatment and increasing the chance to predict them more by the association rules, in this paper we introduced a new partial matching concept in association rules mining. Using this partial matching concept, in our different experiments we found that on average 60% to 80% missing values are imputed with association rules while rest are impute using k- nearest neighbor approach.

## 2 Related Work and their Limitations

Due to the frequent occurrence of missing values in training observations, imputation or prediction of the missing data has always remained at the center of attention of KDD and data mining research community. Imputation is a term that denotes the procedure to replace the missing values by considering the relationships present in the observations. In [6] imputing missing values using prediction model is proposed. To impute missing values of attribute $X$, first of all a prediction model is constructed by considering the attribute $X$ as a class label and other attributes as input to prediction model. Once a prediction model is constructed, then it is utilized for predicting missing values of attribute $X$. The main advantages of imputing missing values using this approach are that, this method is very useful when strong attribute relationship exists in the training data, and secondly it is a very fast and efficient method as compared to k-nearest neighbor approach. The imputation processing time depends only on the construction of prediction model, once a prediction model is constructed, then the missing values are imputed in a constant time. On the other hand, the main drawbacks of this approach are that, if there is no relationship exists among one or more attributes in the dataset and the attributes with missing data, then the prediction model will not be suitable to estimate the missing value. The second drawback of this approach is that, the predicted values are usually more accurate than the true values.

In [4] another important imputation technique based on k-nearest neighbor is used to impute missing values for both discrete and continuous value attributes. It uses majority voting for categorical attributes and mean value for continuous value attributes. The main advantage of this technique it that, it does not require any predictive model for missing values imputation of an attribute. The major drawbacks of this approach are the choice of using exact distance function, considering all attributes when attempting to retrieve the similar type of examples, and searching through all the dataset for finding the same type of instances, require a large processing time for missing values imputation in preprocessing step.

## 3 Partial Matching of Association Rules in HMiT

In [3] the association rules are added in the fired set by comparing the known attributes values of the missing value observation with the frequent itemset or antecedent part of the association rules. If all the known attributes of the missing value observation matches with all the items of antecedent part of association rule X, then X is added in the fired rule set F, otherwise rejected. The main flaw of using this approach is that, in our experimental results on different real datasets, we found that on average 40% to 70% of missing values are imputed using association rules while rest are imputed using k-nearest neighbor approach. Which not only decrease the missing values imputation accuracy but also increases the processing time. To further increase the accuracy of missing values imputation, we propose a partial matching concept between the known attributes value of missing value observation and the items of antecedent part of association rules. Using this partial matching concept an association rule R is added in the fired rule set F, if at least p% of known attributes value of X matches with the items of antecedent part of rule R, where p% is the partial matching threshold. This partial matching concept can be considered from the following definition.

**Definition 1: -** Let X denotes the missing value observation and S is the set of association rules, where A denotes the antecedent part and B denotes the consequent part of association rules. An association rule in S comes under the concept of partial matching, if at least p% attributes value of X matches the items of A.

## 4 Hybrid Missing Values Imputation Technique (HMiT) with Partial Matching

Let $I = \{i_1 \ldots i_n\}$ be the set of n distinct items. Let *TDB* represents the training dataset, where each record *t* has a unique identifier in the *TDB*, and contains set of items such that $t_{items} \subset I$. An association rule is an expression *A -> B*, where *A* and *B* are items $A \subset I$, $B \subset I$, and $A \cap B = \phi$. Here, *A* is called antecedent of rule and *B* is called consequent of rule.

---

**HybridImputation (Support *min_sup*, Confidence *min_conf*)**

1. Generate frequent itemset (*FI*) on given *min_sup*.
2. Generate association rules (*AR*) using *FI* on given *min_conf*.
3. for each training observation *X*, which contains at least on attribute value missing.
4.    for each rule *R* in association rules set (*AR*).
5.      compare the antecedent part of *R* with the known attribute values of *X*.
6.        if antecedent part of *R* matches at least p% of *X* {known attribute values}.
7.          add *R* to fired set *F*
8.    if $F \neq \phi$
9.      if the missing value of *X* is discrete then use *Mod* in set *F* to impute missing value.
10.      if the missing value of *X* is continuous then take *Median* in set *F* to impute missing value.
11.    else
12.      impute missing value using k-nearest neighbor approach.

---

**Fig. 1.** Pseudo code of partial matching-HMiT missing values imputation technique

The main technique of partial matching-HMiT is divided into two phases- (a) firstly the missing values are impute on the basis of partial matching association rules by comparing the antecedent part of rules with the known attributes values of missing value observation, (b) For the case, when there is no association rule exist or fired against any missing value (i.e. no relations exist between the attributes of other observations with the attributes of missing value), then the missing values are imputed using the imputation technique based on k-nearest neighbor approach [4]. The main reason why we are using k-nearest approach as a hybrid combination is that, it is considered to be more robust against noise or in the case when the relationship between observations of dataset is very small.

At the start of imputation process a set of strong (with good support and confidence) association rules are created on the basis of given training dataset with support and confidence threshold. Once association rules are created, the partial matching-HMiT utilizes them for missing values imputation. For each observation *X* with missing values, association rules are fired by comparing the known attributes values of *X* with the antecedent part of association rules one by one. If the know attributes values of *X* matches at least p% items of the antecedent part of association rule *R*, then *R* is added in the fired set *F*, otherwise rejected. Once all association rules are checked against the missing value of *X*, then the set *F* is considered for imputation. If the set *F* is non empty, then median and mod of the consequent part of the rules in set *F* is used for missing value imputation in case of numeric and discrete attributes. For the case, when the set *F* is empty, then the missing value of *X* is imputed using k-nearest neighbor approach. The pseudo code of our partial matching-HMiT is described in Figure 2. Lines from 1 to 2 create the association rules on the basis of given support and confidence threshold. Lines from 4 to 7 first compare the antecedent part of all association rules with the known attributes values of missing value observation *X*. If the known attributes values of *X* are subset of any rule *R*, then *R* is added in the set *F*. If the set *F* is non empty, then the missing values are imputed on the basis of consequent part of fired association rules in Line 9 and 10, otherwise the missing values are impute using k-nearest neighbor approach in Line 12.

## 5 Experiments Results

To evaluate the performance of partial match-HMiT we performed our experimental results on different benchmark datasets available at [7], and check its accuracy against partial matching-HMiT, full matching-HMiT and k-nearest neighbor approach by considering various factors such as support vs. accuracy, support vs. number of missing values imputed, confidence vs. accuracy, confidence vs. number of missing values imputed and random missing values vs. accuracy. The brief introduction of each dataset is described in Table 1. To validate the effectiveness of partial matching-HMiT, we add random missing values in each of our experimental dataset. For performance reasons, we use Ramp [2] algorithm for frequent itemset mining and association rules generation. All the code of partial matching-HMiT is written in Visual C++ 6.0, and the experiments are performed on 1.8 GHz machine with main memory of size 256

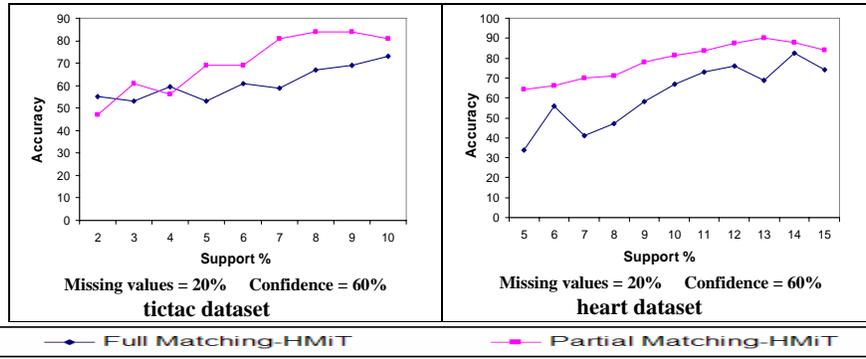

**Fig. 2.** Effect of Missing Values Imputation Accuracy on Support Threshold

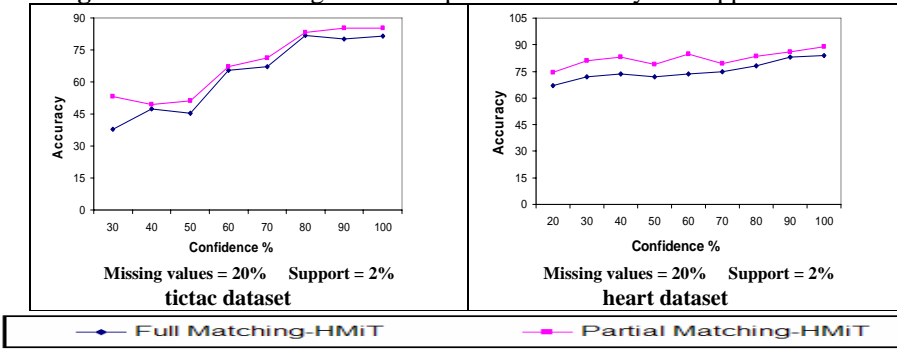

**Fig. 3.** Effect of Missing Values Imputation Accuracy on Confidence Threshold

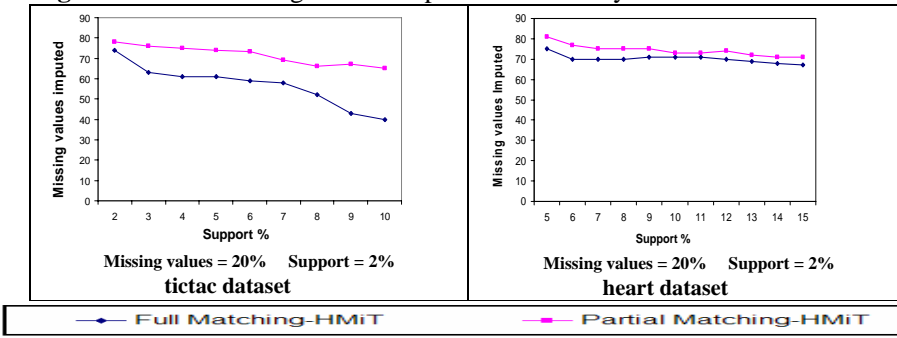

**Fig. 4.** Number of Missing values imputed using Full Matching-HMiT and Partial Matching-HMiT

MB, running Windows XP 2005.

**Table 1.** Details of our experimental datasets

| Datasets | Instances | Total Attributes | Classes |
|---|---|---|---|
| heart | 270 | 13 | 2 |
| tictac | 958 | 10 | 2 |
| Car Evaluation | 1728 | 6 | 4 |
| CRX | 690 | 16 | 2 |

### 5.1 Effect of Support and Confidence Threshold on Missing Values Imputation Accuracy

The effect of support threshold on imputation accuracy is shown in Figure 2 with partial matching-HMiT and full matching-HMiT. The results of Figure 2 are obtained by fixing the confidence threshold as 60% and random missing values are inserted by the percentage of 20%, while the support threshold was varied between 2% to 10%. As clear from the Figure 2 results partial matching-HMiT approach outperforms the other two approaches on almost all levels of support threshold, this due to that more missing values are imputed using partial matching association rules and full matching association rules.

The effect of confidence threshold on missing values imputation accuracy is shown in the Figure 3 with three different imputation techniques (partial matching-HMiT and full matching-HMiT). We obtain the Figure 3 results by fixing the support threshold as 2% and insert random missing values

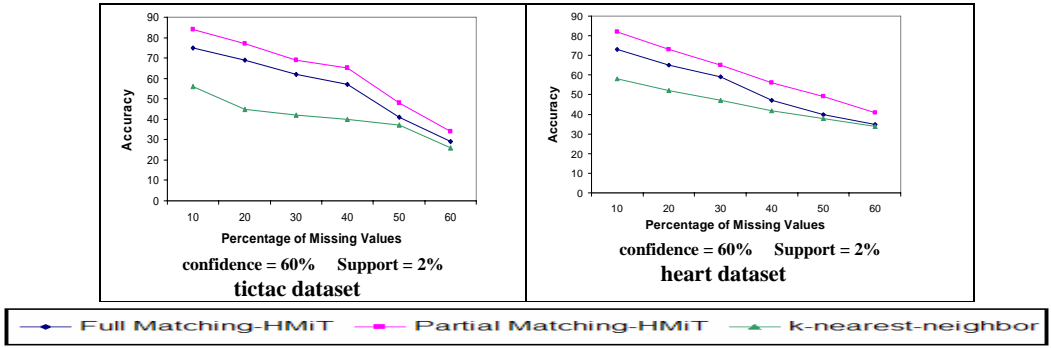

**Fig. 5.** Effect of Percentage of Missing Values on Imputation Accuracy

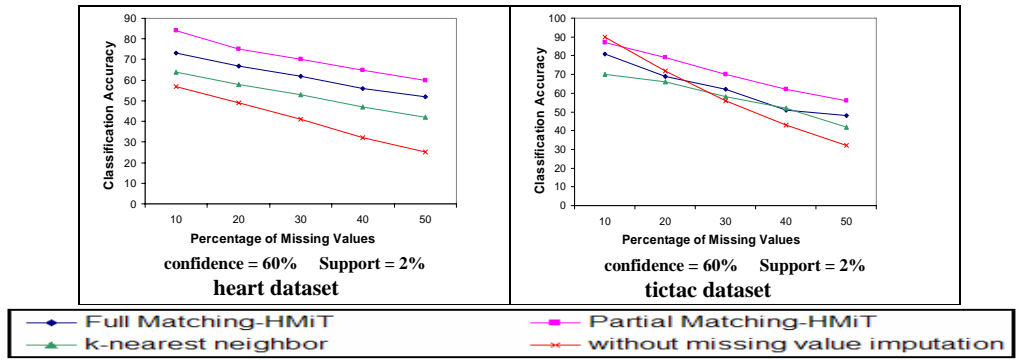

**Fig. 6.** Effect of Missing Values Imputation on Classification Accuracy

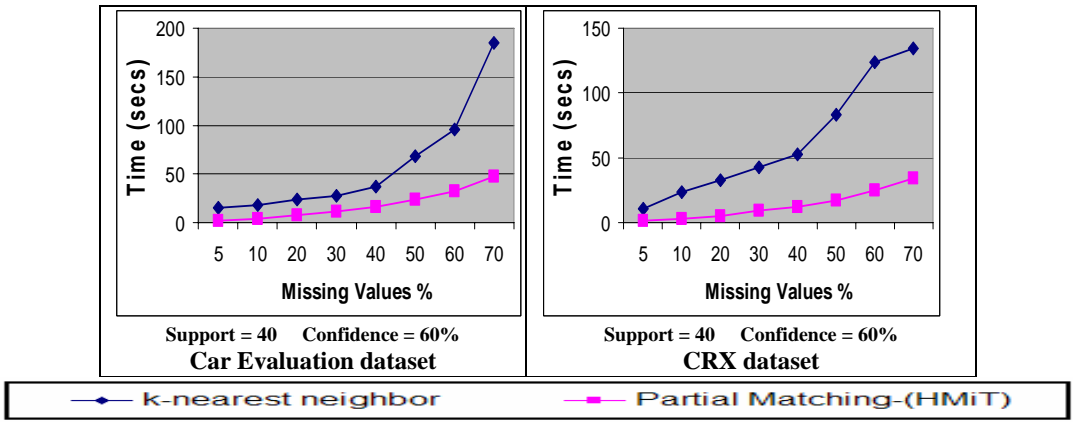

**Fig. 7.** Performance analysis of *HMiT* and k-nearest-neighbor with different missing values level

with 20%, while the confidence threshold was varied from 20% to 100%.

For clear understanding, we exclude the accuracy effect of k-nearest neighbor from Figure 3 results. By looking the results, it is clear that as the confidence threshold increases the accuracy of predicting correctly missing values with partial matching-HMiT as compared to full matching-HMIT also increases. For higher level of confidence only strong rules are generated and they more accurately predict the missing values, but in case of less confidence more weak or exceptional rules are generated which results in very less accuracy. From our experiments we suggest that confidence threshold between 60% to 70% generates good results.

### 5.2 Percentage of Missing Values Imputed by Full Matching-HMiT and Partial Matching-HMiT and Effect of

### Percentage of Missing Values on Imputation Accuracy

In Figure 4 we show the percentage of missing values imputed by two hybrid missing values imputation techniques (partial matching association rules and full matching association rules). The results are obtained by fixing the confidence threshold as 60% and random missing values are inserted with 20%, while the support threshold was varied from 2% to 10%. As clear from Figure 4 results, partial matching-HMiT imputes more missing values than full matching-HMiT on almost all levels of support threshold which shows the effectiveness of this approach.

The effect of percentage of missing values on imputation accuracy is shown in Figure 5 with partial matching-HMiT, full matching-HMiT and k-nearest neighbor approach. The results in Figure 5 are showing that as the level of percentage of missing values increases the accuracy of predicting missing values decreases. We obtain the Figure 5 results by fixing the support and confidence thresholds as 2% and 60% respectively and nearest neighbor size as 10. The reason behind why use nearest neighbor size = 10, is descried in [1]. From the results it is clear, that partial matching-HMiT generates good results as compared to full matching-HMiT and k-nearest neighbor approach on almost all levels of random missing values.

### 5.3 Effect of Missing Values Imputation on Classification Accuracy

In order to perform missing values imputation on missing values data, we should be very clear whether the missing values imputation will increase or decrease the learning model accuracy. To answer this question, we select supervised datasets and perform complete classification accuracy tests with missing values imputation techniques such as k-nearest neighbor, partial matching-HMiT and full matching-HMiT. For experimental purposes we used the weak 3.4 tool available at [8] and select the J 4.8 classification algorithm. Our results in Figure 6 show that as the percentage of missing values increases the classification accuracy without missing values imputation decreases as compared to with missing values imputation. On the other hand partial matching-HMiT gives the best classification accuracy results on very high percentage of missing values. The results of Figure 7 are showing that partial matching-HMiT performs better in term of processing time as compared to k-nearest neighbor approach.

## 6 Conclusion

Missing value imputation is a complex problem in KDD and data mining tasks and in recent years lot of techniques are proposed of handling this task. The main limitation of previous techniques is that, they depends upon single approach and do not combine multiple approaches, that's why they are less accurate. In this paper we introduced a novel partial matching concept in association rules for better treatment of missing values. We evaluate the effectiveness of this partial matching concept; we add it in our hybrid missing values imputation techniques (partial matching-HMiT) and performed detailed experimental results on various benchmark datasets. Our results suggest that missing values imputation using partial matching-HMiT not only improve the missing values imputation accuracy and processing time, but it also increase the classification accuracy, which shows the effectiveness of our concept.